\documentclass[10pt,conference]{IEEEtran}

\pdfoutput=1
\usepackage[caption=false]{subfig}
\usepackage[singlelinecheck=false]{caption} 

\usepackage{hyperref}
\usepackage{graphicx}
\usepackage{comment}
\usepackage{amsmath}
\usepackage{amssymb}
\usepackage{amsfonts}
\usepackage{bbm}
\usepackage{braket}
\usepackage{booktabs}
\usepackage{verbatim}
\usepackage{tabularx}
\usepackage{cite}
\usepackage{ulem}
\usepackage{color, colortbl}
\usepackage{multirow}
\definecolor{LightCyan}{rgb}{1,0.5,0.5}
\hypersetup{
  colorlinks   = true, 
  urlcolor     = blue, 
  linkcolor    = blue, 
  citecolor   = blue 
}

\usepackage{color}
\usepackage{tikz}

\newcommand{\be}{\begin{equation}}
\newcommand{\ee}{\end{equation}}

\newcommand{\citefuture}[1]{}

\begin{document}

	\title{
	{Stochastic Approach For Simulating Quantum Noise Using Tensor Networks}
	}

\author{\IEEEauthorblockN{
William Berquist \IEEEauthorrefmark{1}\IEEEauthorrefmark{2}\IEEEauthorrefmark{5},
Danylo Lykov\IEEEauthorrefmark{1}\IEEEauthorrefmark{3}\IEEEauthorrefmark{5}, Minzhao Liu\IEEEauthorrefmark{1}\IEEEauthorrefmark{4}\IEEEauthorrefmark{5}, and Yuri Alexeev\IEEEauthorrefmark{1}}
\IEEEauthorblockA{\IEEEauthorrefmark{1}Computational Science Division, Argonne National Laboratory, Lemont, IL 60439}
\IEEEauthorblockA{\IEEEauthorrefmark{2}Department of Computer Science, University of Houston, Houston, TX 77004}
\IEEEauthorblockA{\IEEEauthorrefmark{3}Department of Computer Science, University of Chicago, Chicago, IL 60637}
\IEEEauthorblockA{\IEEEauthorrefmark{4}Department of Physics, University of Chicago, Chicago, IL 60637}
\IEEEauthorblockA{\IEEEauthorrefmark{5} These authors contributed equally to the paper}
}

	\date{Started 6/3/21}

	\maketitle

\begin{abstract}
Noisy quantum simulation is challenging since one has to take into account the stochastic nature of the process.
The dominating method for it is the density matrix approach.
In this paper, we evaluate conditions for which this method is inferior to a substantially simpler way of simulation.
Our approach uses stochastic ensembles of quantum circuits, where random Kraus operators are applied to original 
quantum gates to represent random errors for modeling quantum channels. We show that our stochastic simulation error is relatively low, even for large numbers of qubits. We implemented this approach as a part of the QTensor package.
While usual density matrix simulations on average hardware are challenging at $n>15$, we show that for up to $n\lesssim 30$, it is possible to run embarrassingly parallel simulations with $<1\%$ error. By using the tensor slicing technique, we can simulate up to 100 qubit QAOA circuits with high depth using supercomputers.

\end{abstract}

\section{Introduction}

Quantum information science (QIS) has a great potential to speed up certain computing problems like combinatorial optimization and quantum simulations \cite{alexeev2019quantumworkshop}. The development of fast and resource-efficient quantum simulators to classically simulate quantum circuits is the key to the advancement of the QIS field. Currently, we are in the Noisy Intermediate-Scale Quantum (NISQ) era of quantum computing. Therefore, it is particularly important that noisy quantum simulators are developed in order to help develop, test, and verify the quantum algorithms we hope to use. 

There are many types of quantum simulators \cite{wu2019full, wu2018amplitude, wu2018memory, quac,markov2008simulating, pednault2017breaking, boixo2017simulation, lykov2021large}, and tensor network simulators have shown the state-of-the-art performance. However, when it comes to simulating quantum computers with noise, the current very common approach is to use the density matrix formalism. This approach allows one to obtain an exact noisy state with a single sample, but it has a memory cost that scales at $4^n$, where $n$ is the number of qubits. We use a tensor network representation and apply noise stochastically, generating an approximate noisy state. This has a much lower memory cost that scales at $2^n$, but it requires many samples and therefore has a much higher computation cost. Despite this higher computation cost, the lower memory cost allows us to simulate larger quantum systems that are intractable using the density matrix formalism. Thus, we effectively traded memory requirements for more demanding computational requirements. This tradeoff is especially attractive for running large-scale simulations on supercomputers.

We have implemented our stochastic noise model in the tensor network simulator QTensor \cite{qtensor,lykov2021large,lykov2021importance}, which is specifically designed to run in parallel mode on GPU supercomputers at scale. Our eventual goal is to run large-scale quantum circuit simulations on Argonne's supercomputers Polaris and Aurora.

We have tested our implementation of the stochastic quantum simulator in QTensor against the density matrix simulator in the Qiskit package. It has been tested by running a variety of Quantum  Approximate  Optimization  Algorithm (QAOA)~\cite{farhi2014quantum} quantum circuits.

\section{Related Work}

The $O(4^n)$ complexity of simulating noisy quantum circuits and open quantum systems, in general, using density matrices, has sparked decades of development of various algorithms. Approximating the full-density matrices with lower-rank alternatives is the common theme behind all of the approaches.

Tensor network methods such as matrix product states (MPS) represent wavefunctions as factorized tensors, which were originally proposed to simulate many-body quantum systems with local interactions. In systems such as the transverse field 1D Ising model, interactions between quantum spins are limited to the nearest neighbor. The overall statevector is represented as a chain of tensors, each corresponding to a single spin. Each tensor has open bonds (exposed and unconnected to anything) that correspond to the actual physical Hilbert space of spins, as well as closed bonds (connected between tensors) that represent an internal (virtual) degree of freedom. To perfectly represent an exponentially large Hilbert space, the number of virtual bonds between each pair of tensors (bond-dimension) has to grow with the number of qubits, leading to an exponential simulation cost. However, truncating the tensor by limiting the bond-dimension can lead to approximate results with tunable simulation costs. Such truncations are performed with singular value decomposition (SVD). For noisy simulations of 1D systems, statevectors need to be generalized to density matrices. As a result, MPS are generalized to matrix product operators (MPOs), with bonds representing the normal and dual indices.

Other techniques such as time-evolving block decimation (TEBD) for noisy time dynamics simulations, density matrix renormalization group (DMRG) for ground state search in 1D systems, projected entangled pair states (PEPS) for 2D systems, tree tensor networks, and multi-scale entangle renormalization ansatz (MERA) for highly entangled 1D states with global order parameters, also use various representations of quantum states that are low rank. A recent approach for weakly noisy simulations projects the density matrix onto ensembles of pure states, which is more memory efficient.

One potentially interesting area of research is to use of deep-learning techniques for the probabilistic simulation of quantum circuits. It is an exact formulation of quantum dynamics via factorized generalized measurements, which maps quantum states to probability distributions with the advantage that local unitary dynamics and quantum channels map to local quasi-stochastic matrices. Using this framework, quantum circuits that build Greenberger-Horne-Zeilinger states and linear graph states of up to 60 qubits have been demonstrated \cite{carrasquilla2021probabilistic}. Another interesting recent work is \cite{nguyen_tensor_2021}, where a tensor network is constructed using the density matrix instead of statevector. This approach, however, requires significantly more memory with a growing number of qubits and thus can be impractical for systems of $n>15$.

\section{Methodology}
\subsection{Introduction to QTensor}

Using tensor network representation of quantum circuits allows efficiently simulate many-qubit circuits with small depth. In this formalism, each gate operation is represented by a tensor, where indices correspond to each input and output state. If two gates act on the same qubit, the corresponding tensors share an index. The whole circuit is a collection of tensors that are connected by shared indices, which is called a tensor network. Evaluation of the probability amplitude requires summation over the shared indices through the process of tensor network contraction. To this end, a sequence of contracted indices is ordered. For each index, a list (bucket) of corresponding tensors is formed, which is called \textit{bucket index}. Each bucket is a collection of tensors that share the same bucket index. Buckets are contracted one by one using a tensor multiplication library. It is done by summing over the bucket index, and the resulting tensor is then appended to the appropriate bucket.

The memory requirement of tensor network contraction is high (exponential) and corresponds to the largest number of indices of a single tensor encountered during contraction called the \textit{contraction width}. As a result, contraction along the qubit direction rather than the time direction allows a reduction of contraction width and simulation costs. It is especially efficient for the simulation of shallow quantum circuits. In this work, we used the Argonne-developed tensor network simulator QTensor~\cite{qtensor}. It is developed for running large-scale quantum circuit simulations using modern GPU-based supercomputers. It has been used to perform the largest QAOA simulations in the world. QTensor utilizes state-of-the-art heuristic tensor contraction order optimizers (third-party and own custom optimizers), which substantially reduce the simulation cost by minimizing the contraction width of the contraction sequence. We used a number of techniques to speed up simulations. For more information, see the following papers: \cite{lykov2021importance, lykov2021large, shutski2019adaptive}.

\subsection{QTensor Backends}

QTensor has support for a few tensor contraction libraries (backends) for contracting tensors efficiently:
\begin{itemize}
    \item Numpy: a CPU-optimized option.
    \item PyTorch: a CPU and GPU option with backpropagation capabilities, which is especially useful for optimization simulations such as QAOA and neural network simulations.
    \item CuPy: a GPU option.
    \item cuTensor: a dedicated GPU library developed by NVidia for efficient tensor contractions.
\end{itemize}

The optimal choice of a backend(s) depends on the target hardware and the particular task~\cite{qtensor_gpu}. Moreover, since these backends are constantly evolving, the optimal choice may change.



\subsection{Index Slicing}
In a high-performance computing environment, the possibility of parallelization must be exploited to achieve low time-to-solution. Although tensor contractions are highly parallel operations that can be done on a GPU since elements of tensors can be processed in parallel, it is hard to naturally utilize multiple machines at this level of parallelism. On a whole quantum circuit level, the tensor network can be contracted in parts by fixing a value of some tensor indices. However, this necessarily changes the nature of the contraction and the contraction width. The step-dependent slicing algorithm~\cite{lykov2021large} we provide within QTensor is a heuristic algorithm that distributes contraction operations of different slices of the tensor network to a different machine in parallel that also balances the contraction width changes.

\subsection{Parallelism Hierarchy}
QTensor implementation of stochastic noise has three levels of parallelism. The first level of parallelism is sample parallelism, where each sampled circuit is simulated independently. Thus, the first level can be trivially parallelized. Depending on the treewidth of the circuits, we use different strategies. For circuits with low contraction width (meaning that the memory requirement is low), multiple circuits can fit into a single GPU. In this case, a single GPU can simulate a batch of circuits in parallel, and multiple GPUs/nodes can be used at the same time. For circuits with larger contraction treewidth, multiple GPUs must be used to simulate a single circuit since the intermediate tensor will not fit in the memory of a single GPU. Multiple GPU nodes need to be used to simulate multiple circuits in parallel using the tensor-slicing technique described in the previous section.

The second level of parallelism is circuit parallelism. As discussed in the index slicing section, a single large circuit can be contracted in parallel on multiple GPUs/nodes by dividing the graph into multiple parts for parallel contraction.

The third level of parallelism is tensor parallelism. This is simply the parallelism allowed by GPUs when processing independent tensor elements as opposed to CPUs.

Overall, with the three levels of parallelism in mind, we hope to run our noisy simulation on the Polaris supercomputer, which is especially suitable for this task for its thousands of latest-generation GPUs available as well as the state-of-the-art communication fabric.

\subsection{Quantum Approximate Optimization Algorithm}
QAOA is hailed as one potential approach to achieving quantum advantage on NISQ devices. This technique aims to solve an optimization problem, namely the MaxCut problem. Given a graph, we need to find the best way to split the nodes into two groups such that the maximum number of edge connections between nodes are severed by the grouping. The brute force search algorithm that tries each grouping will have to explore an exponentially large space, and this is intractable. QAOA encodes a potential solution in the basis state of the quantum Hilbert space. Each node has a corresponding qubit, and a Hamiltonian is constructed such that every edge that connects two nodes corresponds to a spin-spin interaction term in the Hamiltonian. Finding the optimal solution corresponds to finding the bit string wavefunction that minimizes this Hamiltonian or the ground state. Since a quantum circuit can represent an exponentially large number of basis states as a superposition, it is believed that with the appropriate state preparation and optimization schemes, QAOA can find the lowest energy bit string with high probability.

Recent work shows that for the good performance of QAOA on NISQ devices, circuit ansatze with shallow depths tend to perform better. This is partially explained by the fact that shallower circuits accumulate less noise. This fact is particularly favorable since QTensor is especially well-suited for simulating low-depth circuits. Our effort to develop a noisy version of the simulator can have a significant impact on the search for an efficient algorithm under realistic hardware constraints.

\subsection{Modeling Noise}
\begin{figure}
    \centering
\includegraphics[width=0.9\linewidth]{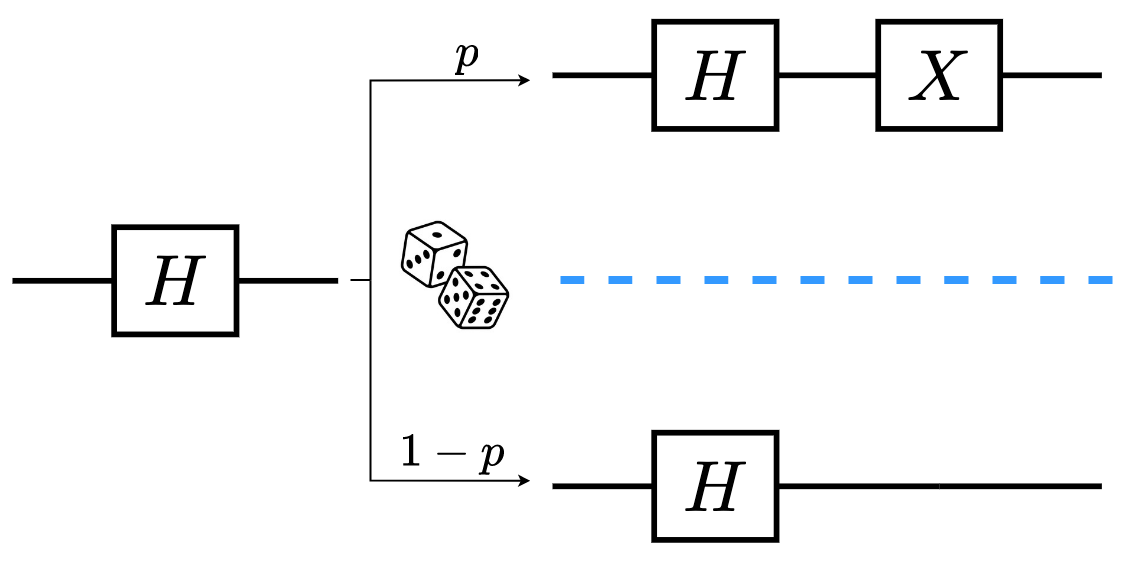}
    \caption{An example of stochastic bit-flip noise on a single qubit gate. When an $H$ gate is applied to a circuit, it has a probability $p$ of also applying an $X$ gate, which is the bit-flip noise. Otherwise, only an $H$ gate is applied as an ideal, noiseless gate with probability $1 - p$.}
    \label{fig:stochastic_noise_example}
\end{figure}

 The general idea behind noise models is that whenever an operation is done on a circuit, there is some probability \(1-p\) of just that operation happening, and there is some probability \(p\) that operation plus an unwanted operation occurring. The unwanted operation is the noise. An example is shown in Figure \ref{fig:stochastic_noise_example}.

\begin{figure}
    \centering
\includegraphics[width=0.9\linewidth]{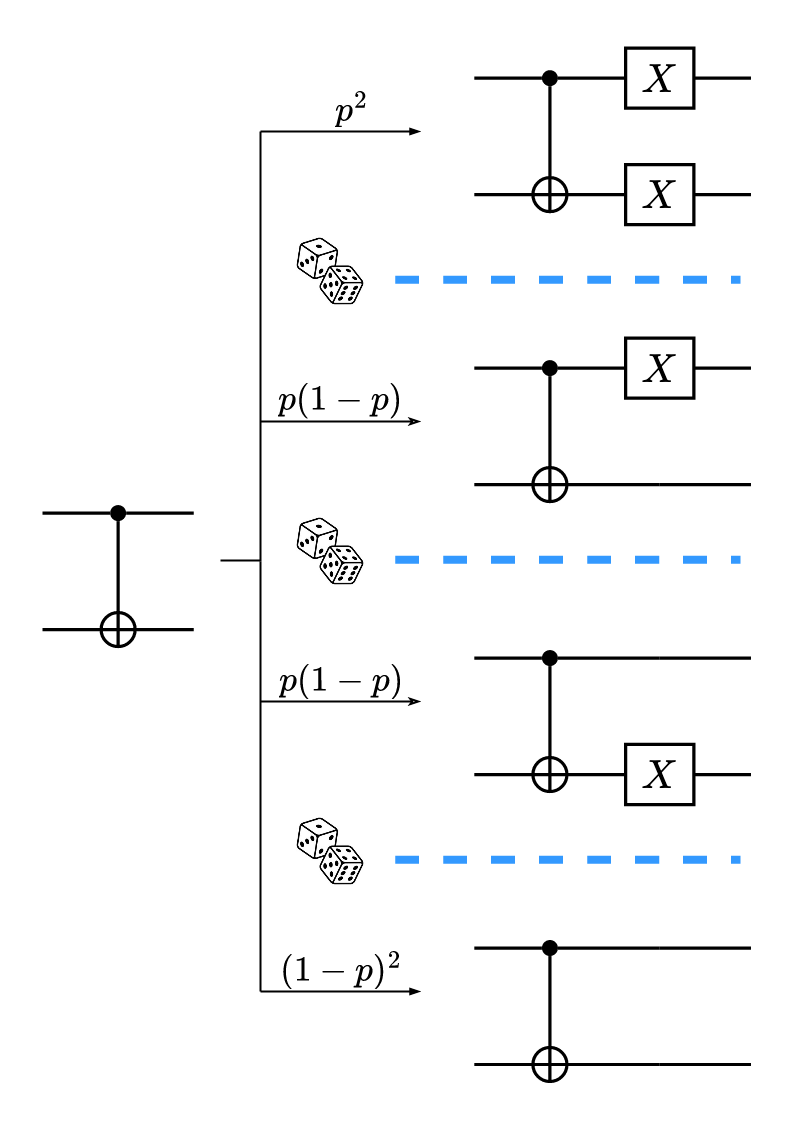}
    \caption{An example of stochastic bit-flip noise, with probability $p$ of being applied on a single gate, applied to a two-qubit gate. When a $cX$ gate is applied to a circuit, each qubit has some probability of also applying an $X$ gate, which is the bit-flip noise. With probability $p(1 -p)$, only the target qubit or only the control qubit will have bit-flip noise applied. With probability $p^2$, both qubits will get the bit-flip noise. Finally, with probability $(1 - p)^2$, no noise will be applied. }
    \label{fig:stochastic_noise_2qubit_gat_example}
\end{figure}
We can express errors in the density matrix formalism using the operator-sum representation \cite{nielsen00}. An open quantum system can be modeled as 

\begin{equation}
    \label{eqn:op_sum}
    \mathcal{E}(\rho) = \sum\nolimits_{j} K_{j} \rho K_{j}^{\dagger}
\end{equation}
where \(\mathcal{E}\) is a linear map called a \textit{channel}. Any evolution in quantum mechanics is called a channel - both unitary and irreversible - and channels convert systems from one state to another. Each \(K_j\) is called a \textit{Kraus operator}. The Kraus operators for a bit-flip channel are given by 

\begin{equation}
    K_0 = \sqrt{p}X, \quad K_1 = \sqrt{1 - p}I
    \nonumber
\end{equation}
and the Kraus operators for a depolarizing channel on a single qubit are given by 

\begin{equation}
    K_{0}=\sqrt{1-\frac{3 \lambda}{4}}I, \,\,
    K_{1}=\sqrt{\frac{\lambda}{4}} X, \,\,
    K_{2}=\sqrt{\frac{\lambda}{4}} Y, \,\,
    K_{3}=\sqrt{\frac{\lambda}{4}} Z
    \nonumber
\end{equation}

While \(p\) in the bit-flip channel directly refers to a probability, \(\lambda\) in the depolarizing channel is a parameter that only corresponds to a probability. We use square roots because each \(K_j\) is multiplied by its complex-conjugate transpose in Equation \ref{eqn:op_sum}.

\subsection{Stochastic Noise Implementation in QTensor}

There are several steps to simulating stochastic noise. First, a noise model is created, which contains a list of all of the noise channels that the circuit will have. Each channel is associated with a particular gate or gate that will be applied in the circuit. Then an ideal, noiseless circuit is created. Finally, a function \verb+simulate_batch_ensemble()+ is called, which has the ideal circuit, the noise model, and the number of circuits in the ensemble $K$ as arguments. 

Every ideal circuit in the ensemble is recreated in the exact order it was originally created, except with noise. First, gate $i$ from the ideal circuit is added to the noisy circuit. Then there is a check to see if that $i$ is in the noise model. If it is not, the next gate from the ideal circuit is added. If gate $i$ is in the noise model, then we begin to add noise based on the channels associated with that gate. For each noise channel associated with the gate, we generate a uniform random number $0 \leq u \leq 1$ and use $u$ to pick a Kraus operator from the channel. We then apply the Pauli operator (or operators if it is a multi-qubit channel) associated with the Kraus operator to the noisy circuit. For example, if the Kraus operator picked from a bit-flip channel is $\sqrt{p}X$, then we apply the Pauli operator $X$ to the noisy circuit. The application of the Pauli is the noise. After all of the noise channels for gate $i$ are added, then gate $i + 1$ from the ideal circuit is added, and we do the checks again. 

After every gate from the ideal circuit has been added to the noisy circuit, we simulate the circuit and obtain a statevector $\psi$. We take the absolute value squared of each element of $\psi$ to obtain a probability density vector $\varphi$

\begin{align}
    \varphi = 
    \sum_j^{2^n} 
    |\psi_j|^2 \,\text{e}_j
    \nonumber
\end{align}

where $\{\text{e}_k\}$ are the standard basis vectors. Note that $\psi$ is normalized in another part of the QTensor package, so the normalization of $\varphi$ is taken care of already. We add $\varphi$ to another vector: the average probability density vector. The average probability density vector keeps track of the results of every noisy circuit simulation from the ensemble. After $K$ simulations of noisy circuits, we normalize the average probability density vector to obtain the approximate noisy state $\sigma_{\text{approx}}$ 

\begin{align}
    \sigma_{\text{approx}} = 
    \sum_j^{K} 
    \frac{\varphi_{j}}{K}
    \nonumber
\end{align}

We can compare $\sigma_{\text{approx}}$ with Qiskit's density matrix simulator by using Qiskit's \verb+AerSimulator()+ backend with the \verb+density_matrix+ method and an equivalent noise model. To ensure we get the exact density matrix, we apply the \verb+.save_density_matrix()+ method to the Qiskit circuit right before measurement. This will give us an exact noisy state in density matrix form, $\rho_{\text{exact}}$ 

Next  we store the diagonal entries of $\rho_{\text{exact}}$ in a vector of $\text{dim}(2^n)$, denoted $\sigma_{\text{exact}}$.

\begin{align}
    \sigma_{\text{exact}} = \sum_j^{2^n} \rho_{jj} \,\text{e}_j 
    \nonumber
\end{align}

where $\rho_{ik}$ are matrix elements of $\rho_{\text{exact}}$.  We do this because the probabilities of $\rho_{\text{exact}}$ are encoded in the diagonal entries. 

Finally, we calculate the error between the states with 
\begin{align}
    \text{Error} 
        &= 1 - F(\sigma_{\text{approx}}, \sigma_{\text{exact}}) \nonumber \\
        &= 1 - \left| \left\langle \sqrt{\sigma_{\text{approx}}}  ,         \sqrt{\sigma_{\text{exact}}}
        \right\rangle 
        \right|^{2}
\end{align}
where $F(\cdot \,,  \cdot)$ is the fidelity between the states and $\langle \cdot \, , \cdot \rangle$ is the inner product function. We take the square roots of each vector because the inner product should be performed on probability amplitude vectors, not probability density vectors. 

\begin{figure}
    \centering    \includegraphics[width=\linewidth]{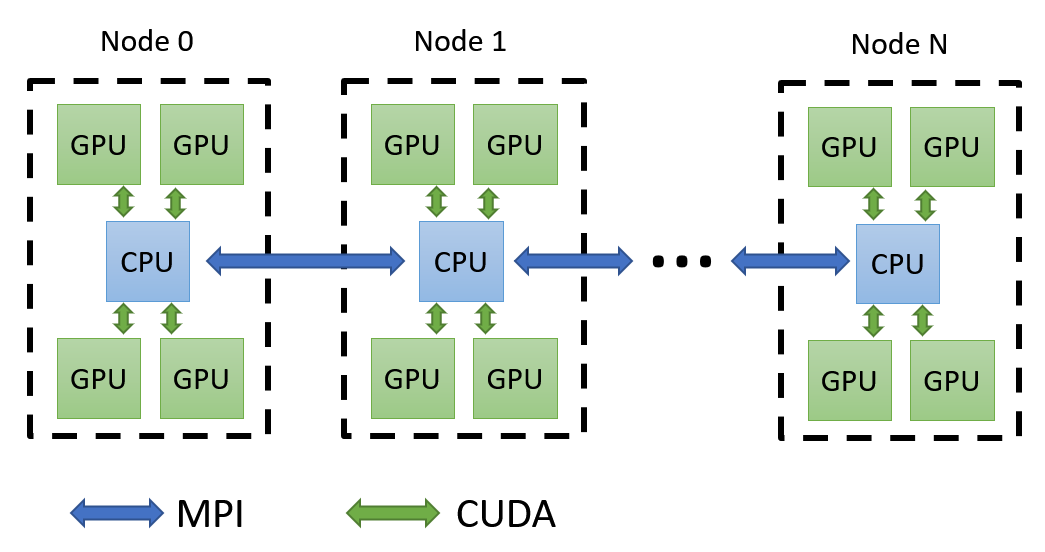}
    \caption{
    The architecture of the Polaris supercomputer at the node level and programming models used to parallelize the QTensor package. 
    }
    \label{fig:architecture}
\end{figure}

\subsection{Computational Resources}

All presented calculations have been obtained with a computer that has a 2.60Ghz Intel i7-9850H 6-core CPU with 16 GB DDR4 RAM, a 512 GB SSD, and an Intel UHD Graphics 630. 

\begin{figure}
    \centering
\includegraphics[width=0.7\linewidth]{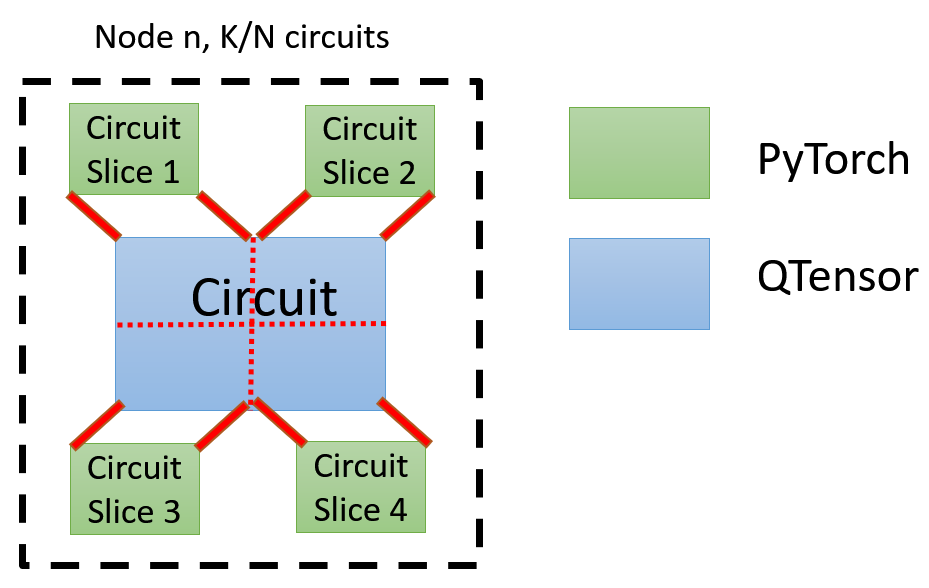}
    \caption{Circuit mapping to a single Polaris node. The circuits are sliced into four parts, which are mapped to GPUs. All calculations are done using QTensor on CPUs and Pytorch on GPUs.}
    \label{fig:slicing}
\end{figure}

Our eventual goal is to run the accurate large-scale noisy quantum simulation using QTensor on Argonne's supercomputers Polaris and Aurora. Polaris is a 560-node HPE Apollo 6500 Gen 10+ based system. Each node has a single 2.8 GHz AMD EPYC Milan 7543P 32-core CPU with 512 GB of DDR4 RAM and four Nvidia A100 GPUs, a pair of local 1.6TB of SSDs in RAID0 for the users use, and a pair of slingshot network adapters. The architecture of Polaris is shown in Figure \ref{fig:architecture} at the node level. To decrease the memory requirements to store circuits in memory, we sliced circuits to decrease contraction width. This algorithm is described in our other paper \cite{lykov2021large}. The mapping of circuits is shown in Figure \ref{fig:slicing}.

\section{Results}

We tested the error between the noisy quantum states generated by the QTensor and Qiskit using many different QAOA circuits. Each ensemble contained between 10 and 1,780 circuits. Each circuit had a depth of $p = 2$, degree $d = 4$, and between $3$ and $13$ qubits. Values for $\gamma$ and $\beta$ were fixed. We added depolarizing noise on all of the gates for our noise model. We used $\lambda_1 = 0.001$ for single-qubit gates, and $\lambda_2 = 0.004$ for two-qubit gates. We chose depolarizing error for two reasons. One is because this is a very common error that is experienced on current quantum computers today. And two, it is one of the worst types of gate error, and it has the largest impact on fidelity. 

At first glance, these values for $\lambda_1$ and $\lambda_2$ may seem small, as they correspond to error rates an order of magnitude lower than those experienced for single- and two-qubit gates on current IBM superconducting devices. However, the values for $\lambda_1$ and $\lambda_2$ had two constraints. First, if we chose values for $\lambda_1$ and $\lambda_2$  that corresponded to error rates experienced today and used those with a QAOA algorithm with the parameters listed above, the noise would overpower the QAOA algorithm, leaving us with a state that is indistinguishable from a uniform distribution state. That is, our final probability amplitude distribution would have the fidelity of $> 0.99$ with the uniform distribution state. Second, if we chose values for  $\lambda_1$ and $\lambda_2$ that were too small, then the final probability amplitude distribution would have a fidelity $> 0.99$ with the exact, noiseless state. 

Our choice of  $\lambda_1$ and $\lambda_2$ met both of these constraints for the parameters $p$, $d$, and number of qubits $n$. The average fidelity between $\sigma_{\text{approx}}$ and the uniform distribution state was $0.938$, and the average fidelity between $\sigma_{\text{approx}}$ and the noiseless state was $0.959$. 

We fixed $\gamma$ and $\beta$ for a similar reason to our choice of $\lambda_1$ and $\lambda_2$. If we used randomized values for $\gamma$ and $\beta$, some of our final states would end with a fidelity $> 0.99$ to the uniform distribution state, some would end with a fidelity $> 0.99$ to the noiseless state, while others could be very far away from those state: e.g. $< 0.8$ fidelity from the uniform or noiseless state. By fixing $\gamma$ and $\beta$ to angles from~\cite{fixed_angle} we removed that dependence. 

\begin{figure}
    \centering
\includegraphics[width=\linewidth]{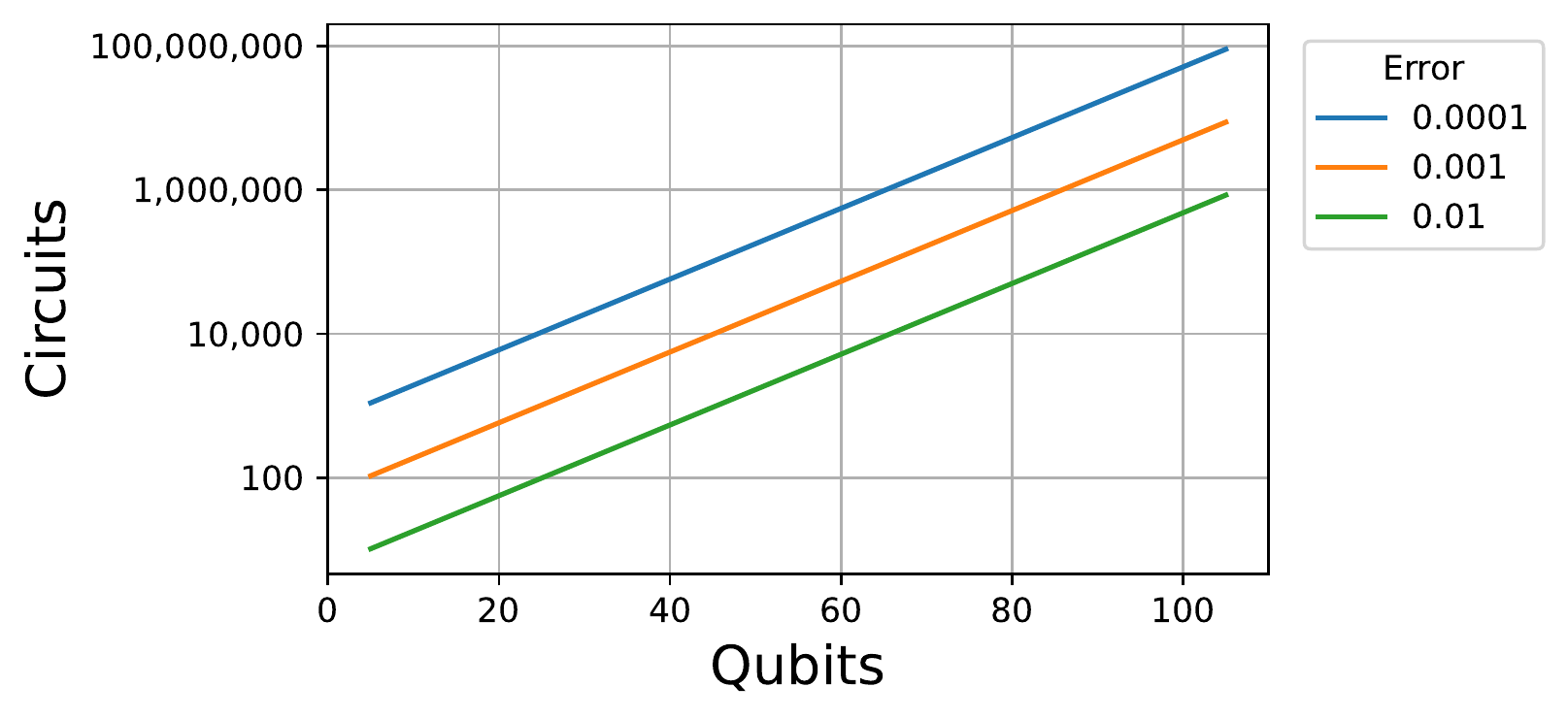}
    \caption{Breakdown of the number of circuits needed to achieve a given error. This is what our model predicts based on Equation \ref{eqn:error_vs_circuits}. This demonstrates that for a fixed number of qubits, a lower error rate requires more circuits in the ensemble. Additionally, for a fixed error rate, more qubits also require more circuits.}
    \label{fig:error_vs_circuits}
\end{figure}

\begin{figure}
    \centering
\includegraphics[width=\linewidth]{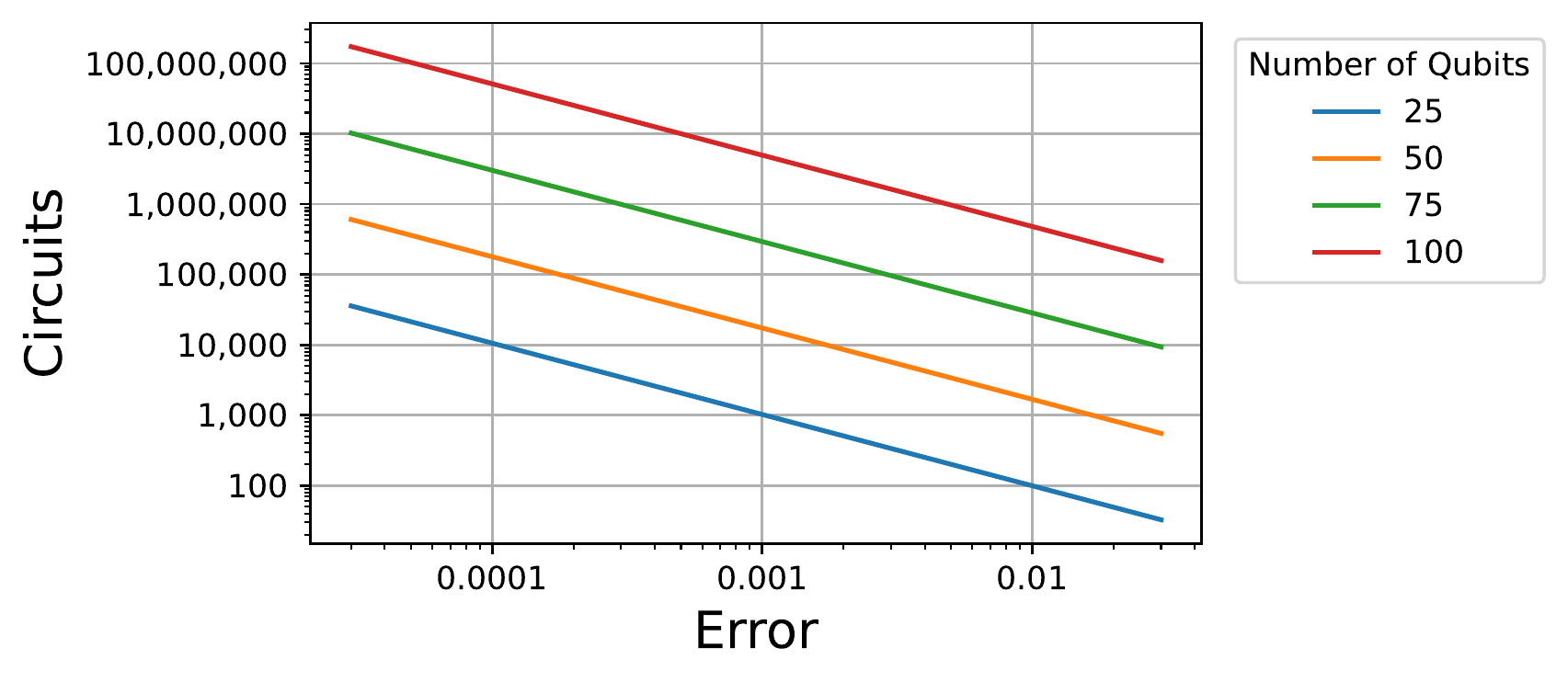}
    \caption{Breakdown of the number of circuits needed to simulate a certain number qubits, for different fixed error rates.}
    \label{fig:qubits_vs_circuits}
\end{figure}

What we found was that as we increased the number of qubits in our simulation but kept the number of circuits in the ensemble fixed, the error increased exponentially. If we kept the number of qubits fixed but increased the number of circuits, the error would decrease. We quantified these results using multiple linear regression, giving us 

\begin{equation}
\label{eqn:error_vs_circuits}
\text{Error} = 
\alpha\,\text{exp}
\big( \delta\;\text{Qubits} -\mu\;\text{ln(Circuits)} \big),
\end{equation}
where $\alpha =  0.05737$, $\delta = 0.11164$, $\mu = 0.98682$ and with $R^2 = 0.996$. Figure \ref{fig:error_vs_circuits} takes this result and then predicts how many circuits we will need for a given error and the number of qubits. 

The shape of the Error in Equation \ref{eqn:error_vs_circuits} is a function of the growing number of qubits, and the size of the density matrix. The density matrix, which is the minimal representation of a generic noisy quantum state, grows as $O(4^n)$, while each circuit only uses vectors of size $O(2^n)$. Thus, each circuit only represents an exponentially small fraction of full quantum noise information, and therefore for a fixed number of circuits, the error should grow exponentially with $n$. Moreover, the error should go down as a polynomial of the number of circuits $K$. 

We find that for our selection of benchmark circuits $\text{Error}~\propto~\frac{1}{K}$, which may be surprising as stochastic error usually scales as $\frac{1}{\sqrt q}$ for $q$ samples.
Due to the simplicity of our model, the dependence of Error on qubit and circuit counts is well understood. This is why despite fitting on relatively small numbers of qubits $n\leq13$, we can safely extrapolate this to large $n$.
The remarkable result of our preliminary small-scale simulations is that there is no requirement to simulate a large number of circuits to get a reasonable error, as shown in Figure \ref{fig:qubits_vs_circuits}. One can achieve $1\%$ error on up to 100 qubits using the order of only a million independent circuits. These calculations can be done efficiently. Running a large number of independent circuits is a perfect task for supercomputers.

\section{Conclusions}

In this work, we developed, to the best of our knowledge, the first parallel stochastic quantum simulator capable of simulating very large quantum circuits with output close to the exact density matrix simulator. It has been implemented in the Argonne-developed tensor network quantum circuit simulator QTensor. We compared the similarity of approximate noisy states generated by QTensor with exact noisy states generated by IBM’s simulator Qiskit by measuring the fidelity between the density matrices. To demonstrate the accuracy, we simulated QAOA circuits up to 13 qubits and depth $p=2$ and compared them against the density matrix simulator in Qiskit.

We evaluated our performance using QAOA ansatz circuits for a very specific set of circuits (MaxCut, $d=4$ regular graphs). While a more general circuit family is interesting, QAOA circuits serve as a benchmark for a useful quantum algorithm MaxCut, which produces samples biased toward a solution to a combinatorial problem. Another direction of this work is to study a relationship between circuit depth $p$ and Error, as well as error probability $\lambda$ and Error. We also plan to run both density matrix and stochastic noise simulation using the tensor networks on supercomputers and study the time and memory requirements of each method. 

By using approximate stochastic techniques, we significantly reduced memory requirements by increasing computational requirements. For example, to simulate a high-depth noisy circuit with 25 qubits using the density matrix method, 18~petabytes of memory is required, while our method needs only 500 MB. Our stochastic approach will need to run only 1,000 noisy circuits to achieve a $0.001\%$ output error.
The memory requirement for the density matrix simulation can be reduced by using the circuit slicing technique. However, at such scale, it is inefficient, as it will increase the simulation time by at least $2^{25}\approx 33\times10^6$ times.

Our stochastic noise simulator is very well suited to run on supercomputers at scale. It is achieved by running a large number of embarrassingly parallel circuit simulations. Currently, we estimate that we can run up to 35 qubit simulations on supercomputers. But by using the tensor slicing technique, we hope to simulate up to 100 qubit noisy QAOA circuits. It is the subject of our future work.


\section*{Acknowledgments}
William Berquist is supported in part by the U.S. Department of Energy, Office of Science, Office of Workforce Development for Teachers and Scientists (WDTS) under the Science Undergraduate Laboratory Internships Program (SULI). Danylo Lykov and Yuri Alexeev are supported in part by the Defense Advanced Research Projects Agency (DARPA) grant.
This work used the resources of the Argonne Leadership Computing Facility, which is DOE Office of Science User Facility supported under Contract DE-AC02-06CH11357.

\bibliographystyle{IEEEtran}
\bibliography{qis}

\end{document}